\documentclass[runningheads]{llncs}
\usepackage{amsmath,amssymb}
\usepackage{graphicx}
\usepackage{booktabs}
\usepackage{subcaption}
\usepackage{tcolorbox}
\usepackage[numbers,sort&compress]{natbib} % camera-ready: replace with splncs04 + \cite
\usepackage{tikz}
\usetikzlibrary{arrows.meta,positioning,calc}
\usepackage[hidelinks]{hyperref}

\newcommand{\takeaway}[1]{\smallskip\noindent\textit{\textbf{Takeaway.} #1}\smallskip}

\begin{document}

\title{Modeling AI Overreliance
as a Complex Adaptive System}
\titlerunning{ }
\author{Ahana Biswas}
\authorrunning{ }
\institute{School of Computing and Information, University of Pittsburgh,
Pittsburgh PA 15260, USA\\ \email{ahana.biswas@pitt.edu}}
\maketitle

% \begin{abstract}
% Whether AI assistance helps or harms depends less on the model's accuracy than on
% whether users rely on it \emph{appropriately}. We study reliance as a population
% process with an agent-based model in which agents repeatedly solve a task alone, use
% an AI answer, or use-and-verify it; hold a Bayesian belief over AI outcomes; and,
% when networked, learn from peers. Four results give one story: \emph{the environment
% sets the baseline; social learning creates consensus, not overreliance; social proof
% turns reliance into a feedback cascade; and feedback design can prevent collapse.}
% Task difficulty and AI quality set the baseline for both behavioral overreliance and
% calibration regret. A mean-preservation
% theorem---confirmed by a $2{\times}2$ topology$\times$tagging design---shows
% connectivity does not shift aggregate reliance under exchangeable signals; we identify a
% boundary condition under which topology \emph{can} matter (belief transmission via
% opinion dynamics). Visible unverified use suppresses verification, collapsing it, with a
% Brock--Durlauf mean-field tipping analogue; making verification visible or dampening
% social proof reverses the collapse under the tested settings. The upshot frames AI
% reliance as computational social dynamics in which individual learning, peer
% observation, and feedback exposure jointly decide whether a population stays
% calibrated.

% \keywords{Human--AI interaction \and Appropriate reliance \and Agent-based modeling
% \and Social learning \and Information cascades \and Calibrated trust.}
% \end{abstract}

\begin{abstract}
Whether AI assistance helps or harms a population depends less on the model's accuracy
than on whether people rely on it \emph{appropriately}---trusting it when it is right
and checking it when it is not. Yet reliance is usually studied one user at a time. We
model it as a population process: agents repeatedly solve a task alone, accept an AI
answer, or verify it, updating a Bayesian belief about AI quality and, when networked,
learning from peers. Four results form one story. \emph{The environment sets the
baseline}: task difficulty and AI quality fix both overreliance and calibration regret.
\emph{Social learning creates consensus, not overreliance}: a mean-preservation theorem,
confirmed by a $2{\times}2$ topology$\times$tagging design, shows connectivity moves the
aggregate only when influence transmits beliefs. \emph{Social proof turns reliance into
a feedback cascade}: visible unverified use suppresses verification and tips the
population into collective overreliance. \emph{Feedback design can prevent collapse}:
making verification visible or dampening social proof reverses it. Together, the results frame AI reliance as a computational social dynamics problem, where individual learning, peer observation, and feedback exposure jointly shape whether a population remains calibrated.
\end{abstract}

% =====================================================================
\section{Introduction}
% =====================================================================
AI assistants---search copilots, coding assistants, clinical and legal decision
aids---increasingly mediate consequential decisions. Across human factors and human–AI interaction, a recurring finding is that model accuracy alone is not enough; outcomes also depend on \emph{calibrated reliance}, leaning on the system when
it is right and overriding it when it is wrong \citep{leesee2004,parasuraman1997}.
Two failures recur---\emph{overreliance}, accepting wrong AI output
\citep{bucinca2021,bansal2021}, and \emph{underreliance}, discarding useful AI after
seeing it err \citep{dietvorst2015}. The dominant empirical paradigm studies these
for a single individual in a one-shot lab task: a participant sees one AI suggestion
and decides whether to accept it. That design has clarified many individual-level effects, but it leaves out two features that matter in deployment. Reliance is \emph{dynamic}: trust is
learned and revised over many interactions as the user accumulates evidence about
when the tool succeeds \citep{hoffbashir2015}. And it is \emph{social}: users see, or
infer, how peers are using the same tool, and adjust accordingly.

Once users learn over time and observe one another, reliance becomes more than an individual judgment.
When trust is learned from experience \emph{and} from observing others, the
population becomes a coupled dynamical system: each person's reliance is both an
outcome of and an input to everyone else's. Such systems can settle into very
different macro-states---broad calibration, collective overreliance, or shared
skepticism---from nearly identical micro-rules, and can exhibit consensus,
path-dependence, and abrupt tipping that are invisible at the individual level. This shifts the question from whether one user accepts one AI recommendation to how a community of users changes its reliance over time.
Consider a running example: a professional repeatedly using an AI assistant for
information-seeking tasks, who may answer alone, accept the AI's answer, or check it,
and who sees colleagues increasingly accept AI answers unchecked.
Table~\ref{tab:world} maps each model component to this setting.

\begin{table}[h]\centering\small
\caption{Model-to-world mapping.}
\label{tab:world}
\begin{tabular}{ll}
\toprule
Model component & Real-world interpretation\\
\midrule
Solve alone (S) & Rely on own judgment, ignore AI\\
Use AI unverified (A) & Accept AI output without checking\\
Verify (V) & Check sources, search externally, consult another tool/person\\
Trust belief & Learned expectation of AI answer quality\\
Social learning & Updating trust from peers' observed experiences\\
Social proof & Seeing others accept AI makes unverified use feel normal\\
Verification cost & Time, effort, cognitive friction of checking\\
Verification collapse & The population stops checking AI outputs\\
Feedback design & Interface choices that change what peer behavior users see\\
\bottomrule
\end{tabular}
\end{table}

Whether a \emph{population} settles into calibrated reliance, collective
overreliance, or skepticism is a system-level question individual experiments cannot
answer. We treat AI reliance as a case of computational social dynamics: individual
learning rules, peer observation, and feedback exposure jointly determine whether a
population remains calibrated or enters a self-reinforcing state of unverified
reliance.

\paragraph{Contributions.} (i) A micro-founded ABM of human--AI reliance with a
random-utility decision rule, a coherent verification utility from a Dirichlet
belief, and a network channel (\S\ref{sec:model}). (ii) A theoretical analysis: an
oracle benchmark, a mean-preservation theorem with an explicit \emph{scope}, and a
mean-field tipping result (\S\ref{sec:theory}). (iii) Four computational results
(\S\ref{sec:results}): the environment sets the baseline; social learning makes
consensus not overreliance (with a boundary condition for when topology matters);
social proof creates a cascade; and feedback design prevents collapse.

We use the model as a theoretical tool for isolating mechanisms rather than as a point forecast. In this sense, it follows the logic of comparative-statics models in economics and computational social science \citep{brockdurlauf2001,granovetter1978}: the goal is to identify which mechanisms are sufficient to generate a pattern and how that pattern changes as conditions vary. First, the model asks which ingredients can make overreliance collective; in our case, this is a verification-suppressing feedback rather than connectivity alone. Second, it provides \emph{qualitative comparative statics}: the sign and direction of effects that are robust across the swept parameter ranges, such as harder tasks raising overreliance and social proof lowering verification. Third, it clarifies intervention logic by separating levers that change the feedback itself from levers that only change individual costs. The policy implications are therefore conditional and qualitative, and double as testable hypotheses for dynamic, social reliance experiments.

% =====================================================================
\section{Related work}
% =====================================================================
\paragraph{Trust and reliance in AI-assisted decisions.}
\citet{leesee2004} distinguish \emph{trust} (an attitude) from \emph{reliance} (the
behavior it produces) and argue the goal is \emph{calibration}---trust tracking
actual reliability---since both over- and under-trust degrade joint performance
\citep{parasuraman1997,parasuraman2010}. Recent HCI work benchmarks reliance against
a normative baseline rather than counting acceptances, decomposing it into relative
AI- and self-reliance (RAIR/RSR; \citealp{schemmer2023}) and showing that verification
friction and cognitive forcing change behavior \citep{bucinca2021}. Two robust
behavioral regularities anchor our validation: \emph{algorithm aversion}, the sharp
loss of trust after seeing an algorithm err \citep{dietvorst2015}, and \emph{algorithm
appreciation}, the tendency to favor algorithmic advice absent visible error
\citep{logg2019}. This work is individual and
static; it characterizes one person's reliance at one moment, leaving the population
dynamics unmodeled. In contrast, we treat aversion and appreciation as patterns the model should
\emph{recover}, not assume.

\paragraph{Social learning and cascades on networks.}
How beliefs evolve in a connected population is the subject of a large social-learning
literature. \citet{degroot1974} models agents repeatedly averaging toward neighbors;
\citet{golubjackson2010} show such naive updating aggregates information (is ``wise'')
only when no single agent's influence remains large. Rational Bayesian variants can
instead produce \emph{information cascades}, in which private signals stop entering the
public pool and a confident but fragile consensus forms \citep{banerjee1992,bhw1992},
and naive averaging on networks can institutionalize \emph{persuasion bias} by
double-counting well-connected sources \citep{demarzo2003}. \citet{brockdurlauf2001}
show that discrete choice with social interactions admits multiple equilibria above a
coupling threshold---the analytic backbone of our cascade result---which connects to
classic threshold models of collective behavior \citep{granovetter1978}. This
literature, however, rarely takes AI reliance as its object, even though framings of
large models as cultural and social technologies suggest peer dynamics should matter
as much for AI use as for any other adopted technology \citep{farrell2025}. We import
its tools and ask what they imply specifically for verification behavior.

\paragraph{Agent-based modeling methodology.}
Methodologically we follow established ABM practice: structured documentation via the
ODD/ODD+D protocols \citep{grimm2020,muller2013}, pattern-oriented validation to
constrain mechanisms by multiple stylized facts \citep{railsbackgrimm2019}, and global
sensitivity analysis to separate which parameters actually drive outcomes
\citep{thiele2014}.

% =====================================================================
\section{The model}\label{sec:model}
% =====================================================================
We use a minimal model: it includes only the mechanisms needed to study how reliance evolves, so that each result can be linked back to a specific part of the system. A population
of $N$ heterogeneous agents repeatedly faces tasks over $T$ discrete steps. At every
step an agent confronts one task and chooses among three actions that correspond to
the real options a user has---solve it alone (\textbf{S}), accept the AI's answer
without checking (\textbf{A}), or use the AI but verify it (\textbf{V}). The action is
driven by the agent's current \emph{belief} about how reliable the AI is; the agent
then observes an outcome and updates that belief. This single loop---belief
$\to$ action $\to$ outcome $\to$ updated belief---is the engine of the model, and
everything else is structure layered onto it.

We use three modeling choices. (1) The three actions separate
\emph{using} AI from \emph{checking} it, making verification an explicit, costly
behavior whose rise and fall is the central dependent variable---collective
overreliance is, operationally, the population abandoning V for A. (2) Trust is
\emph{learned}, and the key asymmetry---that you learn the truth about the AI only
when you verify, while unverified use yields a noisier, optimism-biased signal---is
built into how beliefs update rather than imposed as a fixed bias; this lets aversion
and appreciation emerge. (3) In the networked version, agents also learn from peers,
and peers' visible behavior can feed back into the action utilities, which is what
makes the system capable of cascading. We describe each component below and give the
complete ODD+D specification, with parameter justifications, in the supplement.
Throughout we write $\theta$ for the choice decisiveness and reserve $\beta$ for the AI
difficulty slope.

\begin{figure}[t]\centering
\begin{tikzpicture}[>=Latex, font=\small,
  box/.style={draw,rounded corners,align=center,minimum height=9mm,inner sep=4pt},
  act/.style={box,fill=blue!6}, st/.style={box,fill=gray!8}]
\node[st] (task) {Task $d_t$,\\ AI quality $q$};
\node[act,right=14mm of task] (eu) {Utilities\\ $U_S,U_A,U_V$};
\node[act,right=14mm of eu] (choice) {Logit choice\\ $P(s)\propto e^{\theta U_s}$};
\node[st,right=14mm of choice] (out) {Outcome\\ (C/P/W)};
\draw[->] (task)--(eu); \draw[->] (eu)--(choice); \draw[->] (choice)--(out);
% red social-proof feedback (above): choice -> utilities, raises U_A
\draw[->,red!75!black,line width=1pt] (choice.north) to[out=90,in=90,looseness=0.85]
  node[midway,above,font=\scriptsize,red!75!black,align=center]
  {social-proof / verification-suppression feedback:\\ visible unverified use raises $U_A$} (eu.north);
% gray learning feedback (below): outcome -> utilities, trust update
\draw[->,gray!65!black] (out.south) to[out=-90,in=-90,looseness=0.6]
  node[midway,below,font=\scriptsize,align=center]
  {trust update (own \& peer experience)} (eu.south);
\end{tikzpicture}
\caption{One step of the model (left to right) with two feedback channels.
The gray loop is ordinary trust learning: outcomes update beliefs, which re-enter
the utilities. The highlighted red loop is the \textbf{verification-suppression
feedback}: visible unverified use by peers raises the utility $U_A$ of using AI
unverified, lowering verification and feeding back into more unverified use. This
loop is what turns an otherwise environment-dominated system into one that can
cascade.}
\label{fig:schematic}
\end{figure}
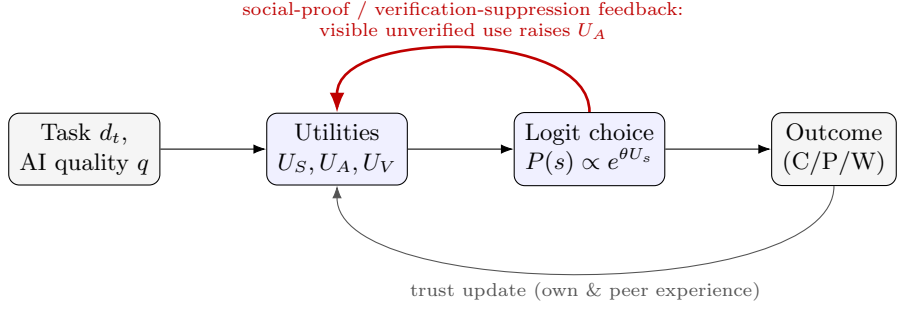

\paragraph{Environment.} Difficulty $d_t\in[0,1]$ is drawn each step; agent $i$ has
ability $a_i$, verification skill $v_i$. AI quality $q\in\{\textsf{H},\textsf{L}\}$
sets $(\alpha,\beta,\gamma,\delta)_q$. To guarantee a valid simplex, ``wrong'' is
conditional on not-correct:
% \begin{equation}\label{eq:ai}
% P(C\mid d)=\sigma(\alpha-\beta d),\
% P(W\mid d)=(1-P(C\mid d))\sigma(\gamma+\delta d),\
% P(P\mid d)=1-P(C)-P(W)\ge0,
% \end{equation}
\begin{equation}\label{eq:ai}
\begin{aligned}
P(C\mid d)&=\sigma(\alpha-\beta d),\\
P(W\mid d)&=(1-P(C\mid d))\,\sigma(\gamma+\delta d),\\
P(P\mid d)&=1-P(C\mid d)-P(W\mid d)\ge0,
\end{aligned}
\end{equation}
with scores $r(C,P,W)=(1,\tfrac12,0)$, solo success
$p^{\mathrm{self}}_i=\sigma(a_0+a_1a_i-a_2d)$, wrong-answer detection
$e_i=\sigma(v_0+v_1v_i-v_2d)$, partial-upgrade $m_i=\sigma(m_0+m_1v_i-m_2d)$.

\paragraph{Beliefs.} Each agent holds a Dirichlet belief $\hat P_i$ over $(C,P,W)$
with trust $\tau_i=\hat P_i(C)+\tfrac12\hat P_i(P)$. Verifying adds $\kappa_V$ to the
true category; unverified use adds $\kappa_U<\kappa_V$ to an optimism-upgraded one;
precision is capped (mean-preserving). The aversion asymmetry is thus emergent from
precision. A scalar delta rule is a robustness variant.

\paragraph{Decision.} Using only its own belief and traits,
% \begin{equation}\label{eq:eu}
% U_S=p^{\mathrm{self}}_i-c_{\mathrm{solve}},\quad
% U_A=\tau_i-c_A,\quad
% U_V=\hat P_i(C)+\hat P_i(P)\tfrac12(1{+}m_i)+\hat P_i(W)e_i p^{\mathrm{self}}_i-c_A-c_{\mathrm{ver}},
% \end{equation}
\begin{equation}\label{eq:eu}
\begin{aligned}
U_S&=p^{\mathrm{self}}_i-c_{\mathrm{solve}},\qquad
U_A=\tau_i-c_A,\\
U_V&=\hat P_i(C)+\hat P_i(P)\tfrac12(1{+}m_i)
     +\hat P_i(W)\,e_i\,p^{\mathrm{self}}_i-c_A-c_{\mathrm{ver}},
\end{aligned}
\end{equation}
($c_{\mathrm{solve}}=\lambda_S d(1-a_i)$, $c_{\mathrm{ver}}=\lambda_V d$). $U_V$
mirrors the realized-outcome rule and uses both verification channels; we treat
$c_{\mathrm{ver}}$ as the total verification burden, including the effort of fallback
correction when a wrong answer is detected (hence no separate $c_{\mathrm{solve}}$ on
that branch). Choice is a
multinomial logit $P(s)\propto\exp(\theta U_s)$ \citep{mcfadden1974,mckelveypalfrey1995}.

\paragraph{Network, cascade, interventions (Model 2).} A fixed ER/BA network
connects agents, who fold visible neighbors' signals into beliefs. We distinguish two
channels: \emph{experience-based} learning, where agents observe neighbors' outcome
signals generated by the shared task environment (so signals are exchangeable), and
\emph{opinion dynamics}, where agents update toward neighbors' current beliefs $\tau_j$
(so biased or high-trust hubs transmit beliefs directly). \emph{Source tagging} credits
each source once, else hubs are over-weighted \citep{demarzo2003}. The \emph{cascade
engine} sets $U_A\!\leftarrow\!U_A+s\,\pi_i$, with $\pi_i$ the influence-weighted
fraction of neighbors who used AI unverified last step. Interventions are symmetric
modifications of this feedback: \emph{verification visibility}
$U_V\!\leftarrow\!U_V+s_V\,\pi_i^V$, with $\pi_i^V$ the influence-weighted fraction who
\emph{verified} last step; \emph{social-proof damping}
$U_A\!\leftarrow\!U_A+\rho\,s\,\pi_i$ with $\rho\in[0,1]$; and \emph{friction reduction}
(lower $\lambda_V$).

\paragraph{Metrics.} Table~\ref{tab:metrics} defines them; regret is primary.

\begin{table}[h]\centering\small
\caption{Metrics. Behavioral proxies vs.\ the normative (oracle) spine.}
\label{tab:metrics}
\begin{tabular}{lll}
\toprule
Term & Definition & Interpretation\\
\midrule
Trust $\tau$ & expected AI score under belief & attitude/belief\\
Behav.\ overreliance & severity-weighted unverified use, $\mathbb 1[A](1-r^{\mathrm{AI}})$ & behavioral proxy\\
Underreliance & solve alone when AI would beat self & behavioral proxy\\
Oracle miscalibration & chosen action $\neq$ oracle action & normative error\\
Regret & utility lost vs.\ oracle & \textbf{primary} metric\\
RAIR / RSR & rely on AI / self when that is better & appropriate reliance\\
\bottomrule
\end{tabular}
\end{table}

% =====================================================================
\section{Theoretical analysis}\label{sec:theory}
% =====================================================================
\begin{definition}[Oracle]\label{def:oracle}
With true expected utilities $U^\star_k$, the oracle is $k^\star=\arg\max_k U^\star_k$;
miscalibration is $\Pr[k\neq k^\star]$ and regret is
$\mathbb E[U^\star_{k^\star}-U^\star_k]\ge0$.
\end{definition}
\begin{proposition}[Oracle reliance]\label{prop:oracle}
A calibrated agent ($\hat P_i$ equal to the true outcome distribution,
$\theta\to\infty$) selects an oracle action $k^\star$ and attains zero expected regret
(up to ties). Behavioral over-/under-reliance are \emph{ex-post} diagnostic proxies
and need \emph{not} vanish under perfect calibration: even an optimal ex-ante action
can yield an unfavorable realized outcome. We therefore use regret as the primary
normative metric and the behavioral proxies as realized-exposure measures.
\end{proposition}
\noindent Indeed, if the oracle plays $A$ when $P(C,P,W)=(0.8,0.1,0.1)$, expected
behavioral overreliance is $0.1(\tfrac12)+0.1(1)=0.15>0$ despite zero regret. Exactly,
\[
\mathbb E[\text{overreliance}]=\underbrace{\Pr[k^\star{=}A]\,\mathbb E[1{-}r^{\mathrm{AI}}\mid k^\star{=}A]}_{\text{irreducible (oracle outcome stochasticity)}}+\underbrace{(\text{signed policy-deviation term})}_{\text{vanishes under oracle-consistent behavior}}
\]
the policy-deviation term being signed (a miscalibrated policy can raise \emph{or} lower
realized overreliance) and vanishing when behavior matches the oracle, which is why
both metrics are reported (full identity in the supplement).
\begin{proposition}[Mean preservation]\label{prop:mean}
Under the DeGroot step $\tau'=(1-\lambda)\tau+\lambda W\tau$ with doubly-stochastic
$W$, the population mean $\bar\tau$ is invariant and the cross-sectional variance is
non-increasing. More generally, if peer signals are exchangeable and the influence
weights are independent of the realized signal values, then
$\mathbb E[\bar\tau']=\mathbb E[\bar\tau]$ for \emph{any} weighting, including
hub-dominated. (Proof in supplement.)
\end{proposition}
\noindent Prop.~\ref{prop:mean} holds exactly for the linear DeGroot trust update; the
Dirichlet observational variant used in the main model follows the same exchangeability
logic (peer pseudo-counts are drawn from the same environment as own experience) and is
verified computationally below.

\begin{tcolorbox}[colback=gray!6,colframe=gray!55,boxrule=0.4pt,left=4pt,right=4pt,
  top=3pt,bottom=3pt]
\small\textbf{Scope of the topology-null result.} Prop.~\ref{prop:mean} is \emph{not}
a claim that networks never matter. It holds when peer signals are exchangeable and
influence is uncorrelated with systematic differences in trust, ability, verification
skill, or exposure. When influence transmits \emph{beliefs} (opinion dynamics) and is
correlated with those traits, topology becomes a mechanism of amplification
(\S\ref{sec:r2}). Exchangeability can also fail through \emph{environment homophily}:
if agents preferentially connect with others facing very different task difficulties or
AI qualities, neighbors' experiences are no longer drawn from a common distribution,
and the null can break even under experience-based learning.
\end{tcolorbox}

\begin{proposition}[Cascade tipping]\label{prop:tip}
Reducing the choice to ``use unverified'' (utility $a_0+s\pi$) vs.\ ``be careful''
($a_0+c$, $c>0$), the mean-field unverified fraction solves $u=\sigma(\theta(su-c))$
(Brock--Durlauf). Real fold points exist iff $\theta s>4$, at
$u_\pm=\tfrac12\big(1\pm\sqrt{1-4/(\theta s)}\big)$; for $c$ strictly between the
corresponding fold values $c_\pm=su_\pm-\tfrac1\theta\log\tfrac{u_\pm}{1-u_\pm}$ there
are three fixed points (two stable, one unstable) and hysteresis, while $c$ outside
$[c_-,c_+]$ gives a unique stable equilibrium. (Proof in supplement.)
\end{proposition}
\noindent Thus $\theta s>4$ is \emph{necessary but not sufficient}: $c$ must also lie in
the fold interval. In the ABM, $\theta=10$ and $s\le0.6$ give $\theta s\le6>4$, yet we
observe no hysteresis (\S\ref{sec:r3}), because heterogeneity, finite-population
stochasticity, and \emph{endogenous} variation in the care advantage $c$ (trust,
verification skill, and difficulty all evolve) keep the system off the fold interval and
smooth the transition. A complementary heterogeneity-attenuation bound---dispersion
$\sigma_c$ multiplies the effective coupling by a factor $\rho(\sigma_c)<1$, so
bistability needs $\theta s>4/\rho$---is derived in the supplement.

% =====================================================================
\section{Experimental design}\label{sec:design}
% =====================================================================
Unless noted, $N=400$, $T=100$, $\theta=10$, and traits are $\mathrm{Beta}(2,2)$;
quantities average the final 10 steps over 20 seeds (18 for dense networked sweeps) with
95\% CIs. The difficulty regimes are $\mathrm{Beta}(2,6)$ (easy) and $\mathrm{Beta}(6,2)$
(hard); the swept ranges are social proof $s\in[0,0.6]$ and learning rate
$\lambda\in[0.1,0.5]$. The full AI, cost, belief, and network parameters, and the Morris
sweep ranges, are listed in the ODD+D supplement; global sensitivity uses Morris
elementary effects.

\paragraph{Continuous AI quality (Fig.~\ref{fig:phase}).} For the phase diagram the
discrete regime $q\in\{\textsf{H},\textsf{L}\}$ is relaxed to a continuous quality
$q\in[0,1]$ that linearly interpolates the AI parameter vector,
$(\alpha,\beta,\gamma,\delta)_q=(1-q)(\cdot)_{\textsf{L}}+q\,(\cdot)_{\textsf{H}}$, so
$q{=}0$ is the Low set and $q{=}1$ the High set.

\paragraph{Intervention settings (Result 4).} All interventions are run in the same
danger zone (Low AI, hard tasks, random$+$tagging) against a harmful baseline of visible
unverified use at $s{=}0.3$. The conditions are: verification visibility $s_V{=}1.0$;
social-proof damping $\rho{=}0.6$ (i.e.\ $U_A\!\mathrel{+}=\!\rho\,s\,\pi_i$); and reduced
friction $\lambda_V{=}0.05$ (from the default $0.20$).

\paragraph{Boundary experiment (Result 2).} Degree-correlated heterogeneity shifts the
\emph{initial trust} of agents by their degree rank: $\tau_0\!\leftarrow\!\tau_0+h_T\cdot
\text{shift}$, where shift runs from $-\tfrac12$ to $+\tfrac12$ across the degree
ordering (clipped to $[0.01,0.99]$). ``High-trust hubs'' use $h_T{=}{+}2.0$, so the
highest-degree agents start most trusting; ``careful hubs'' use $h_T{=}{-}2.0$, the
exact negative, so the highest-degree agents start \emph{least} trusting. The two
conditions differ only in the sign of this initial-trust correlation---neither alters
verification skill---isolating the effect of \emph{which} agents are influential.

\paragraph{Counterfactual outcomes.} behavioral over-/under-reliance and RAIR/RSR
require knowing how the unchosen actions would have fared. For each agent--task we draw
the AI outcome and the agent's solo outcome regardless of the action taken, and evaluate
the realized score of every action; the chosen action drives dynamics, while the full
set benchmarks the behavioral and appropriate-reliance proxies. These counterfactuals
are used only for evaluation, never for agent learning (agents update beliefs solely
from the action they took).

% =====================================================================
\section{Results}\label{sec:results}
% =====================================================================
\subsection{Result 1: The task environment sets the baseline}\label{sec:r1}
\begin{figure}[t]\centering
\includegraphics[width=\linewidth]{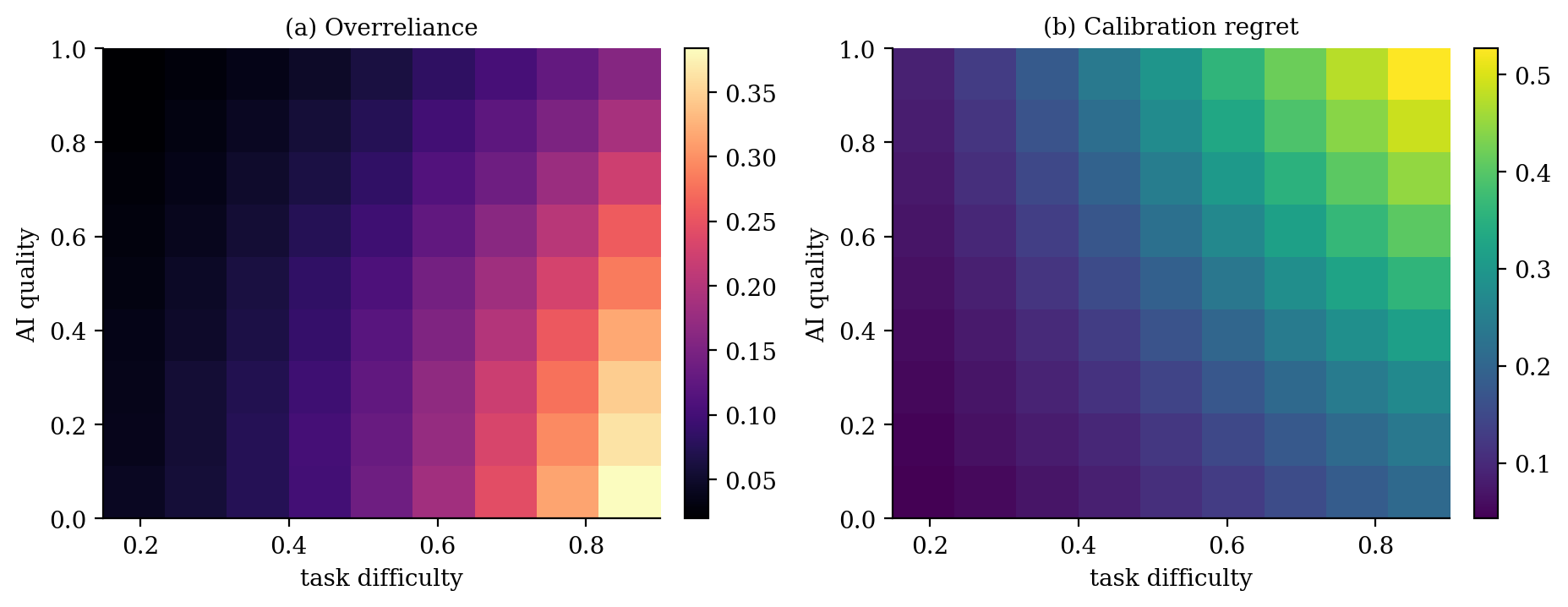}\\[6pt]
\includegraphics[width=0.6\linewidth]{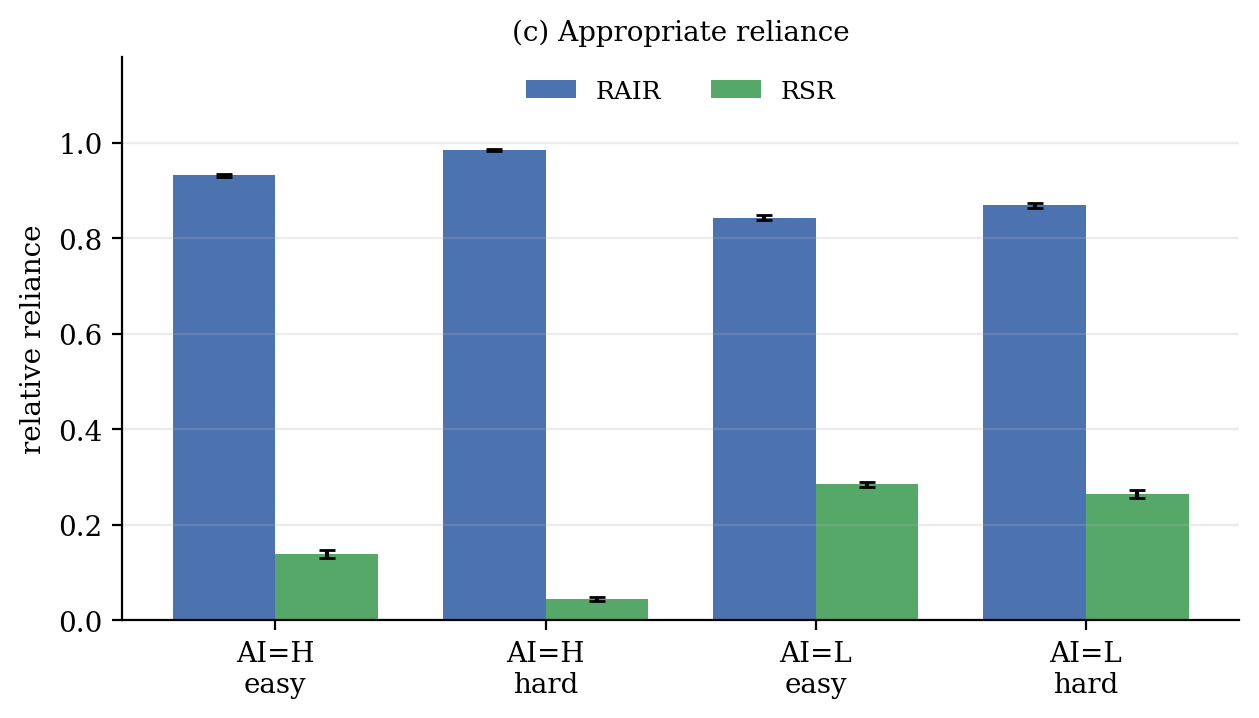}
\caption{\textbf{R1.} Panels \textbf{(a)} overreliance and \textbf{(b)} calibration
regret over task difficulty $\times$ AI quality (the quality axis interpolates
continuously between the Low and High parameter sets; \S\ref{sec:design}): the
environment sets the baseline. \textbf{(c)} High RAIR, low RSR---agents defer to AI but
under-use their own correct judgment (RAIR/RSR: relative AI-/self-reliance on the
disagreement subset).}
\label{fig:phase}
\end{figure}

The phase diagram (Fig.~\ref{fig:phase}) shows overreliance climbing with difficulty
($\approx0.02\to0.38$); at hard tasks AI quality moves it from $0.38$ (poor AI) to
$0.16$ (good AI). A global Morris analysis (Fig.~\ref{fig:sens}a) confirms difficulty
($\mu^*\!=\!0.29$) and AI quality ($0.12$) lead. Regret and behavioral overreliance
diverge informatively (Table~\ref{tab:m1}): high-quality AI on hard tasks has the
\emph{highest regret} ($0.441$) but not the highest overreliance ($0.119$)---agents
over-defer and rarely self-rely (RSR $0.045$); low-quality AI on hard tasks has the
\emph{highest overreliance} ($0.280$) but moderate regret ($0.167$), because the solo
alternative is also poor. The environment sets the baseline; feedback then amplifies
or dampens reliance within it (\S\ref{sec:r3}--\ref{sec:r4}).

\begin{table}[t]\centering\small
\caption{Model 1 final states (20 seeds; CI $\le0.01$). \emph{Regret} is primary.}
\label{tab:m1}
\begin{tabular}{lrrrrrrr}
\toprule
Regime & Overrel. & Underrel. & Trust & Acc. & Regret & RAIR & RSR\\
\midrule
AI=H, easy & 0.032 & 0.023 & 0.904 & 0.932 & 0.148 & 0.932 & 0.139\\
AI=H, hard & 0.119 & 0.009 & 0.830 & 0.843 & \textbf{0.441} & 0.985 & 0.045\\
AI=L, easy & 0.071 & 0.047 & 0.724 & 0.816 & 0.061 & 0.844 & 0.285\\
AI=L, hard & \textbf{0.280} & 0.056 & 0.495 & 0.519 & 0.167 & 0.869 & 0.264\\
\bottomrule
\end{tabular}
\end{table}
\takeaway{In these simulations, task difficulty and AI quality set the baseline level of reliance error, while regret captures the cost of those errors more directly than raw overreliance.}

\subsection{Result 2: Social learning creates consensus, not aggregate overreliance}\label{sec:r2}
\begin{figure}[t]\centering
\begin{subfigure}{\linewidth}\centering\includegraphics[width=0.7\linewidth]{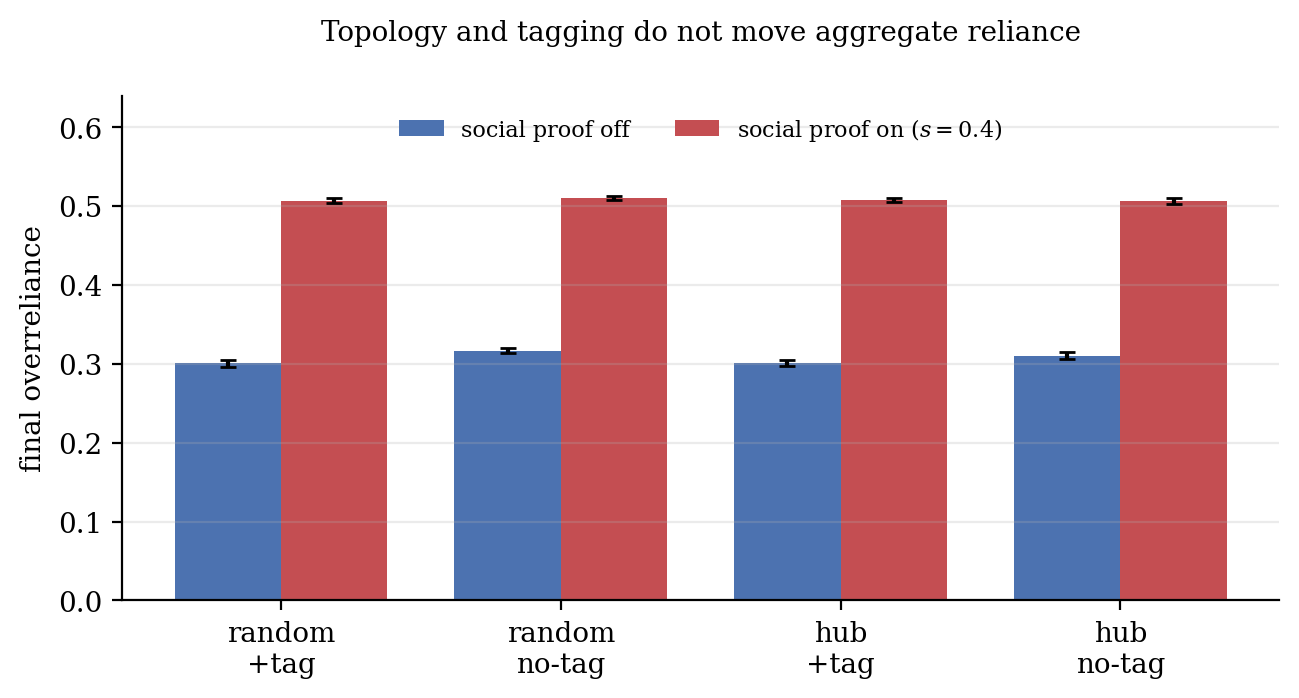}
\caption{Empirical counterpart of Prop.~\ref{prop:mean}.}\end{subfigure}\\[6pt]
\begin{subfigure}{\linewidth}\centering\includegraphics[width=\linewidth]{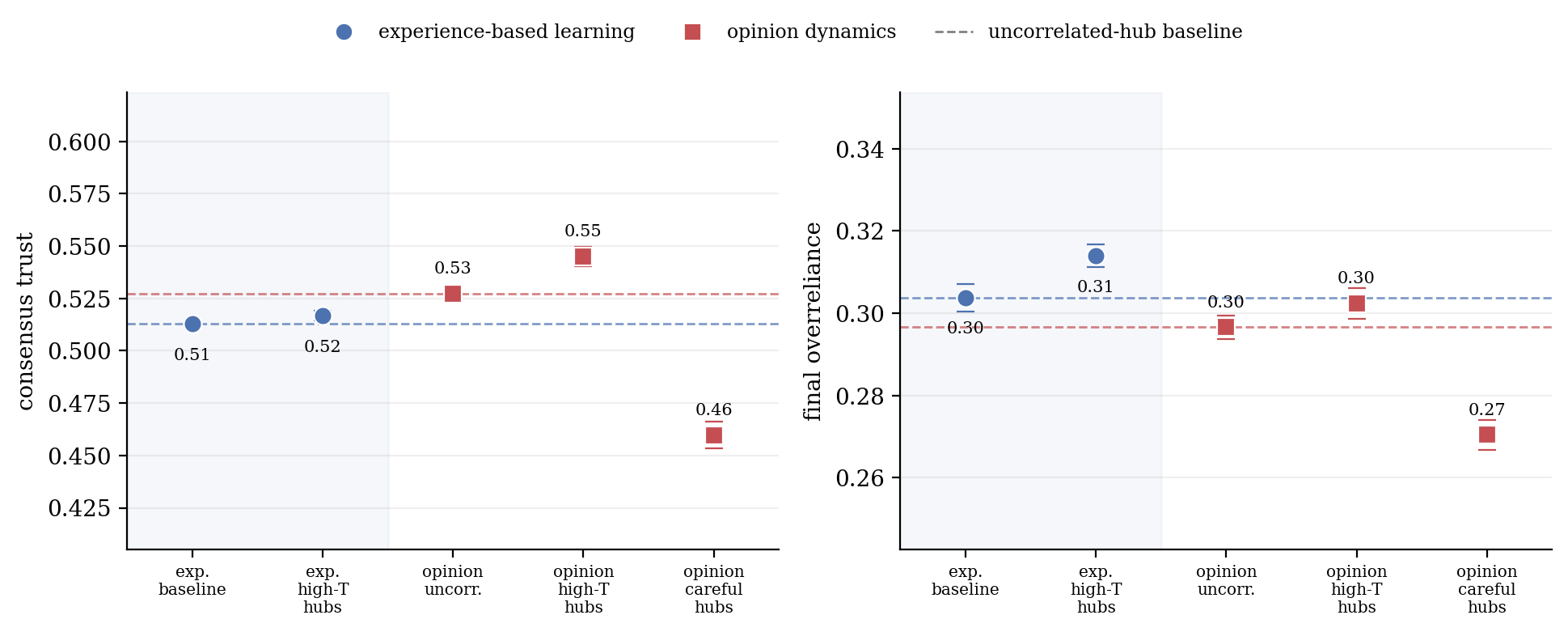}
\caption{When topology matters.}\end{subfigure}
\caption{\textbf{R2.} (a) Connectivity and persuasion bias do not shift aggregate
reliance under exchangeable signals; feedback shifts all conditions together. (b)
Topology matters under belief transmission: experience-based hubs stay near baseline,
while opinion-dynamics hubs shift consensus trust and---for careful hubs---overreliance.}
\label{fig:rob}
\end{figure}

To isolate connectivity from feedback, we cross two network topologies (random ER vs
hub-heavy BA) with source tagging on/off, with and without the social-proof channel
(Fig.~\ref{fig:rob}a). This $2{\times}2$ topology$\times$tagging design is the empirical
counterpart of Prop.~\ref{prop:mean}: with feedback off, all four cells give
overreliance $0.30$--$0.31$ (within CI); with feedback on, all rise together to
$\approx0.51$. Social learning sharply compresses trust dispersion (consensus) while
leaving the aggregate unchanged.

But \emph{when does topology matter?} Fig.~\ref{fig:rob}b locates the boundary.
Under experience-based learning, even extreme high-trust hubs leave the aggregate at
baseline (consensus trust $0.516$ vs.\ $0.513$; overreliance $0.313$ vs.\ $0.303$)---experiential
grounding re-anchors everyone to the environment. Under \emph{opinion dynamics} (peers
transmit beliefs, weak grounding), influential hubs steer the consensus: relative to
uncorrelated hubs (trust $0.530$, overreliance $0.299$), high-trust hubs raise consensus
trust to $0.549$ while careful hubs lower it to $0.455$ and cut final overreliance to
$0.272$. The overreliance shift is muted in the saturated danger zone---raising
already-high trust barely moves behavior---while careful hubs, which move agents across
the verify/use margin, shift it clearly; the consensus-trust shift is unambiguous in
both directions. Thus connectivity can move the aggregate when influence transmits
correlated beliefs, especially when those beliefs move agents across decision margins---a
boundary condition on the null.

\takeaway{Peer learning mainly reduces dispersion in trust. Aggregate reliance shifts only when influential agents transmit correlated beliefs that move others across the verify/use margin.}

\subsection{Result 3: Social proof turns reliance into a feedback cascade}\label{sec:r3}
\begin{figure}[t]\centering
\includegraphics[width=\linewidth]{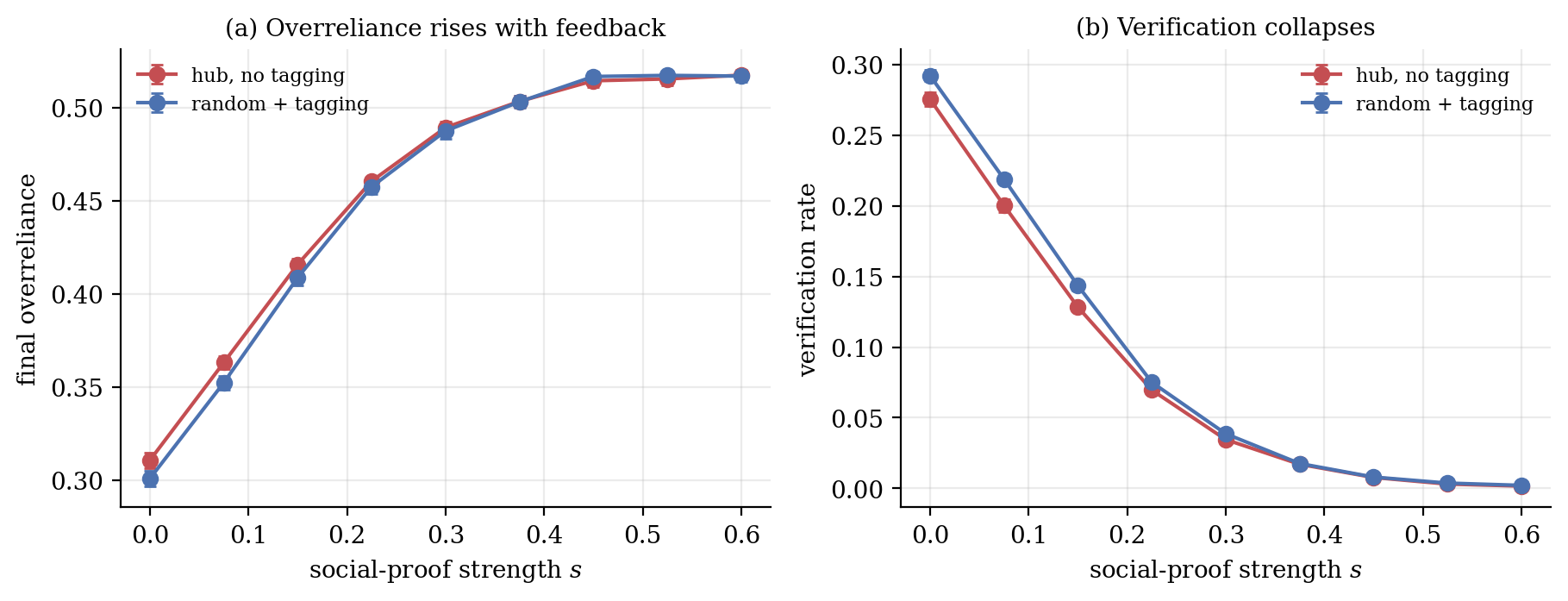}\\[6pt]
\includegraphics[width=\linewidth]{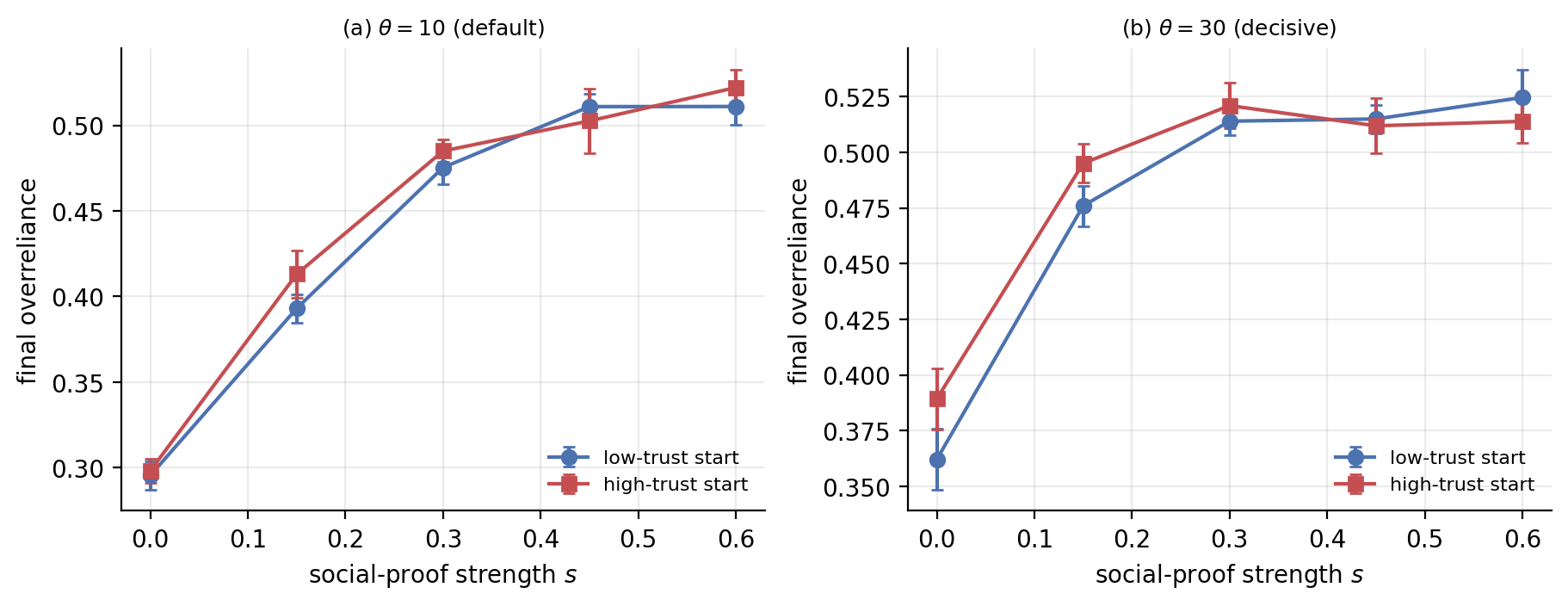}
\caption{\textbf{R3.} \textbf{(a)} overreliance rises and \textbf{(b)} verification
collapses ($<0.01$) as social proof $s$ increases, with hub and random networks
coinciding within CIs. Forward/backward sweeps (low- vs high-trust start) coincide at
\textbf{(c)} $\theta=10$ and \textbf{(d)} $\theta=30$, so the ABM shows a smooth
crossover with no hysteresis; the mean-field saddle-node (Prop.~\ref{prop:tip}) appears
only under strong coupling (supplement).}
\label{fig:cascade}
\end{figure}

Engaging the verification-suppression feedback tips the population from verifying to
collapsed-verification: as $s$ rises $0\to0.6$, verification falls $0.29\to0.002$
($<0.01$ = ``collapse'') and overreliance rises $0.30\to0.52$ (Fig.~\ref{fig:cascade});
hub and random coincide within CIs. Forward/backward sweeps (low- vs high-trust start)
\emph{coincide at every $s$ even at $\theta=30$}: the ABM shows a smooth crossover with
\emph{no hysteresis}; the saddle-node (Prop.~\ref{prop:tip}; mean-field fold in the
supplement) is the strong-coupling analogue. Overreliance here is endogenous---produced
by learning dynamics, not a worse model or worse users.

\takeaway{In the model, collective overreliance appears when peer behavior enters the utility of unverified use and makes verification less attractive.}

\subsection{Result 4: Feedback design can prevent verification collapse}\label{sec:r4}
\begin{figure}[t]\centering
\includegraphics[width=0.74\linewidth]{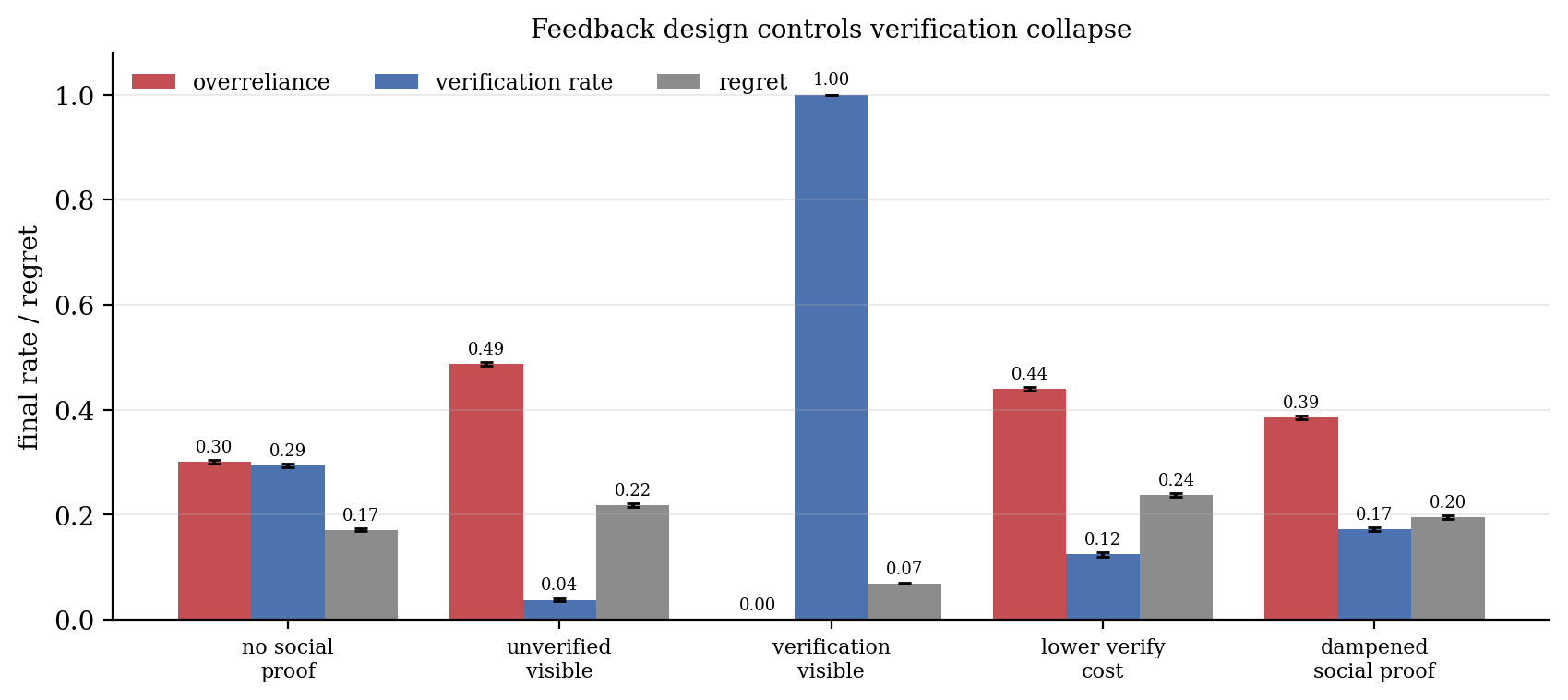}
\caption{\textbf{R4.} Interventions on the feedback channel (Low AI, hard tasks,
harmful baseline at $s{=}0.3$). Making unverified use visible raises overreliance and
collapses verification; strong verification visibility ($s_V{=}1.0$) drives a beneficial
counter-cascade to near-complete verification, while dampening social proof gives a
partial recovery and cheaper verification does not lower regret.}
\label{fig:interv}
\end{figure}

If the cascade comes from feedback, then interventions should be most effective when they alter that feedback channel.
(Fig.~\ref{fig:interv}; intervention strengths in \S\ref{sec:design}). Against the
harmful baseline (unverified use visible: overreliance $0.49$, verification $0.04$,
regret $0.22$), making \emph{verification} visible ($s_V{=}1.0$) triggers a beneficial
counter-cascade---a mirror of the harmful one, with its own tipping point---that, under
this strong setting, moves the population to near-complete verification (overreliance
$0.00$, verification $1.00$) and drives regret down to $0.07$. The effect is
parameter-dependent: it is a counter-cascade above a threshold in $s_V$, not a
guaranteed outcome---for smaller $s_V$ the recovery is partial, with full verification
emerging only above the tipping point (verification rate vs.\ $s_V$ in the supplement).
Dampening social proof ($\rho{=}0.6$) gives a partial recovery
(overreliance $0.39$, verification $0.17$, regret $0.20$). Notably, merely reducing
verification friction ($\lambda_V{=}0.05$) is the weakest lever: it lifts verification
only to $0.12$ and does \emph{not} lower regret ($0.24$), because cheaper checking does
not counter the social pull toward unverified use. Crucially, the interventions that
target the \emph{feedback} (visibility, damping) reduce the primary normative metric,
regret, whereas the one that targets only \emph{cost} does not. The levers are thus the
salience of verification and the structure of exposure---not connectivity.

\takeaway{In this model, interventions that change what users see others doing are more effective than interventions that only lower the private cost of checking.}

\subsection{Pattern recovery, structural plausibility, and sensitivity}
\begin{figure}[t]\centering
\includegraphics[width=0.92\linewidth]{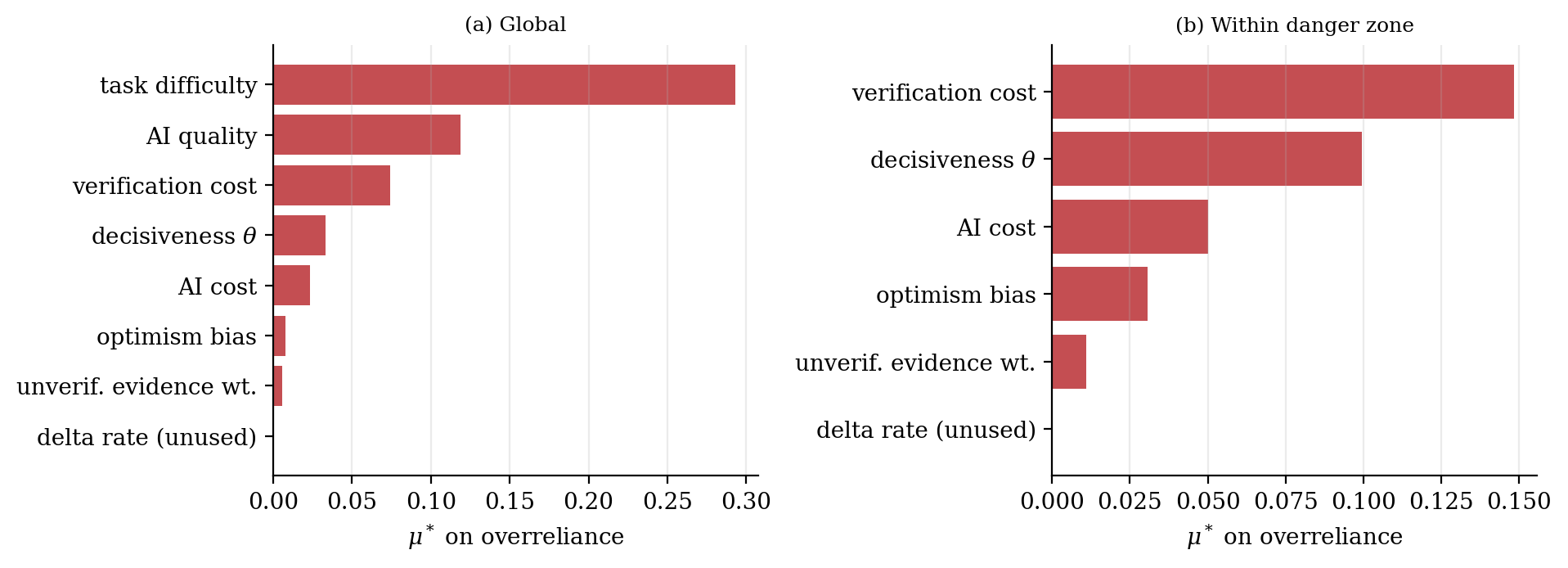}
\caption{Morris $\mu^*$ of \textbf{behavioral overreliance}. (a) Global: difficulty and
AI quality dominate (R1). (b) Within the danger zone: verification cost and decisiveness
dominate.}
\label{fig:sens}
\end{figure}

The model \emph{recovers} three regularities without separately tuning mechanisms for
each: (P1) aversion---poor AI observed via verification yields lower final trust
($0.497$) than unobserved ($0.534$); (P2) appreciation---good AI raises trust
$0.375\to0.871$; (P3) the cascade---feedback raises overreliance $0.303\to0.513$ and
collapses verification $0.292\to0.004$. These are qualitative pattern checks
establishing structural plausibility, not calibration to empirical magnitudes. Morris
sensitivity of \emph{behavioral overreliance} (Fig.~\ref{fig:sens}) reconciles scope:
globally, difficulty and AI quality dominate; within the danger zone, verification cost
($0.15$) and decisiveness ($0.10$) lead, and the unused delta-rate is correctly inert.
(Screening regret instead of overreliance in the danger zone gives the same leading
factor, verification cost, $\mu^*\!=\!0.16$.)

% =====================================================================
\section{Discussion and conclusion}
% =====================================================================
The results point to a simple mechanism. Deployment context sets the baseline, peer learning aligns beliefs, and visible unverified use can erode verification unless the feedback channel is redesigned. Three implications follow. Deployment context dominates, and regret (not raw
overreliance) measures its cost. Connectivity alone is not enough to explain collective overreliance---under exchangeable
signals social learning homogenizes trust without moving its aggregate, and in this
model topology shifts the aggregate only when influence transmits correlated beliefs.
And overreliance is a feedback phenomenon whose levers are the cost and salience of
verification and the structure of exposure: in the settings tested, interfaces that keep
verification cheap and visible, or dampen social proof for unverified use, prevent
collapse.

Consistent with the model's purpose, these implications are \emph{qualitative and
conditional} rather than quantitative forecasts: the model identifies which classes of
intervention can work and why---those that reshape the feedback (making verification
visible, damping social proof) act on the mechanism that produces collective
overreliance, whereas lowering verification cost alone does not, because it leaves the
social pull intact. Equivalently, they are testable hypotheses: a field or longitudinal
experiment that manipulates the visibility of unverified versus verified use should
move population verification rates in the directions the model predicts, and the
$2{\times}2$ and boundary results sharpen what to measure (aggregate reliance is
governed by feedback and by belief-transmitting influential users, not by raw
connectivity).

\paragraph{Limitations and future work.} AI quality is exogenous and stationary, the
network fixed, and verification stylized. The topology null holds for exchangeable
signals; the opinion channel and stake/exposure heterogeneity are where structure bites.
The abrupt-tipping form is a mean-field property; the ABM shows a smooth crossover, so in
practice we expect collective overreliance to build \emph{gradually} as social proof
strengthens rather than to switch on at a sharp threshold---verification erodes
continuously, which is both less dramatic and harder to notice in deployment. Bringing
the model to data would allow estimating a few key quantities from longitudinal traces of
AI-assisted work: per-task verification rates and their dependence on observed peer
behavior (the social-proof strength $s$), the cost and effectiveness of verification,
and how trust updates after verified vs.\ unverified use---quantities measurable in
instrumented tools or controlled multi-session studies.

\paragraph{Reproducibility.} ODD+D supplement (with full proofs) accompany the paper; All codes will be made publicly available on acceptance.

\small
\bibliographystyle{unsrtnat}
\bibliography{main}
\end{document}